\documentclass[aps,prc,twocolumn,superscriptaddress,showpacs]{revtex4-1}
\usepackage{graphicx}
\usepackage{sidecap}
\usepackage{color}
\usepackage{url}

\newcommand{\aNSCL}{\affiliation{National Superconducting Cyclotron Laboratory, Michigan
State University, East Lansing, MI 48824, USA }}

\newcommand{\aMSUphys}{\affiliation{Department of Physics and Astronomy, Michigan State
University, East Lansing, MI 48824, USA}}
\newcommand{\aMSUchem}{ \affiliation{Department of Chemistry, Michigan State University,
East Lansing, MI 48824, USA}}
\newcommand{\aRIKEN}{\affiliation{RIKEN Nishina Center,
RIKEN, Wako-shi, Saitama 351-0198, Japan}}
\newcommand{\aFRIB}{ \affiliation{Facility for Rare Isotope Beams, Michigan State University,
East Lansing, MI 48824, USA}}
\newcommand{\aTAMU}{ \affiliation{Cyclotron Institute, Texas A\&M University, College Station, TX 77843, USA}}

\newcommand{\softsh}[1]{\texttt{#1}}
\newcommand{\soft}[1]{\texttt{#1 }}

\newcommand{\lisepp}{\soft{LISE$^{++}$}}
\newcommand{\liseppsh}{\softsh{LISE$^{++}$}}

\newcommand{\epaxtwo}{\soft{EPAX$\:2.15$}}
\newcommand{\epaxthree}{\soft{EPAX$\:3$}}

\begin{document}

% number in the upper righthand corner of the title page in preprint = mode.
%\preprint{}

\title{Production cross sections from $^{82}$Se fragmentation  as indications  \\ of shell effects in neutron-rich isotopes close to the drip-line}


\author{O.~B.~Tarasov}\email[]{On leave from FLNR/JINR, 141980 Dubna, Russian Federation} \aNSCL
\author{M.~Portillo} \aFRIB
\author{D.~J.~Morrissey} \aNSCL \aMSUchem
\author{A.~M.~Amthor} \aFRIB
\author{L.~Bandura} \aFRIB
\author{T.~Baumann} \aNSCL
\author{D.~Bazin} \aNSCL
\author{J.~S.~Berryman} \aNSCL
\author{B.~A.~Brown} \aNSCL \aMSUphys
\author{G.~Chubarian} \aTAMU
\author{N.~Fukuda} \aRIKEN
\author{A.~Gade} \aNSCL \aMSUphys
\author{T.~N.~Ginter} \aNSCL
\author{M.~Hausmann}\aFRIB
\author{N.~Inabe} \aRIKEN
\author{T.~Kubo} \aRIKEN
\author{J.~Pereira} \aNSCL
\author{B.~M.~Sherrill} \aNSCL  \aMSUphys
\author{A.~Stolz} \aNSCL
\author{C.~Sumithrarachichi} \aNSCL
\author{M.~Thoennessen}\aNSCL \aMSUphys
\author{D.~Weisshaar} \aNSCL

\date{\today}

\begin{abstract}

Production cross sections for neutron-rich nuclei from the
fragmentation of a $^{82}$Se beam at 139~MeV/u were measured. The
longitudinal momentum distributions of 126 neutron-rich isotopes of
elements  \protect{$11\le Z\le 32$} were scanned using an
experimental approach of varying the target thickness. Production
cross sections with beryllium and tungsten targets were determined
for a large number of nuclei including several isotopes first observed in
this work. These are the most neutron-rich nuclides of the elements
\protect{$22\le Z\le 25$} ($^{64}$Ti, $^{67}$V, $^{69}$Cr,
$^{72}$Mn). One event was registered consistent with $^{70}$Cr, and another one with
$^{75}$Fe. The production cross sections are correlated with $Q_{\text{g}}$ systematics to reveal trends in the data.
The results presented here confirm our previous result from a similar measurement using a $^{76}$Ge beam,
and can be explained with a shell model that predicts a subshell closure at $N=34$ around $Z=20$.
This is demonstrated by systematic trends and calculations with the Abrasion-Ablation model that are sensitive to separation
energies.

\end{abstract}

%http://www.aip.org/pacs/pacs2010/individuals/pacs2010_regular_edition/reg20.htm#25
%25.70.Mn 	Projectile and target fragmentation
%27.40.+z 	39 = A = 58
%27.50.+e 	59 = A = 89
%21.60.Cs 	Shell model

%\keywords{}

\maketitle

% body of paper here - Use proper section commands for PR-C

%WWWWWWWWWWWWWWWWWWWWWWWWWWWWWWWWWWWWWWWWWWWWWWWWWWWWWWWWWWWWWWWWWWWWWWWWWWWWWW

\section{Introduction\label{Intro}}

\subsection{Discovery of new nuclei}

The discovery of new nuclei in the proximity of the neutron dripline
provides a stringent test for nuclear mass models, and hence for the
understanding of both the nuclear force and the creation of elements.
Another important aspect of such measurements is that once neutron-rich nuclei are observed and their cross sections for formation
are understood,  investigations to study
the nuclei themselves, such as with decay spectroscopy, can be planned.
Therefore,  obtaining  production rates for the most exotic
nuclei continues to be an important part of the experimental program
at existing and future rare-isotope facilities.

A number of production mechanisms have been used to produce
neutron-rich isotopes for  \protect{$20\le Z\le 28$} ~\cite{OT-PRC09} but, in the last few years, two reaction mechanisms
were the most effective at producing nuclei in this region:
\begin{itemize}
  \item projectile fragmentation -- an experiment with a $^{76}$Ge (132 MeV/u) beam produced 15 new isotopes of  \protect{$17\le Z\le 25$}~\cite{OT-PRL09},
  \item in-flight fission with light targets (Abrasion-Fission) -- an experiment with a $^{238}$U beam~\cite{Ohn-JPSJ10} produced  a large number of isotopes of  \protect{$25\le Z\le 48$} using a Be-target, and several new isotopes with \protect{$46\le Z\le 56$} by Coulomb fission on a heavy target.
\end{itemize}

\begin{SCfigure*}
\includegraphics[width=0.8\textwidth]{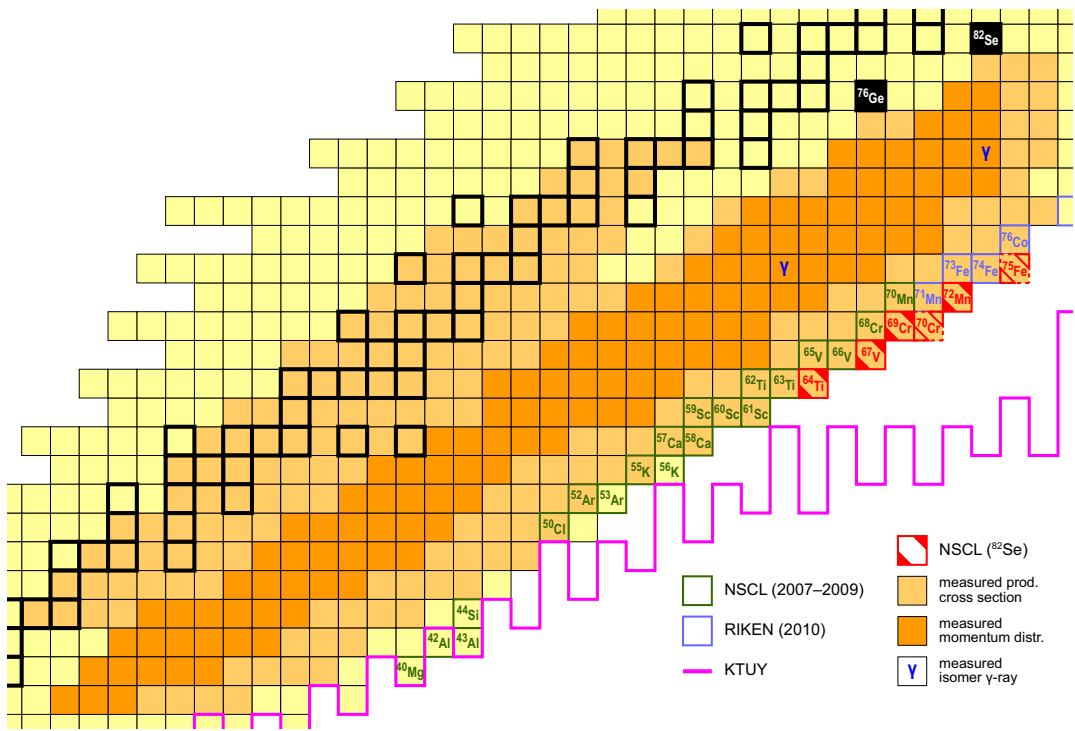}
\caption{(Color online) The region of the nuclear chart investigated
in the   present work. The solid line shows the limit of bound nuclei from the
KTUY mass model~\cite{KTUY-PTP05}. The new isotopes observed for
the first time in the present work are marked by red
squares.\label{chart}}
\end{SCfigure*}

Progress in the production of neutron-rich isotopes was made possible
by the increase of primary beam intensities, new beam development at the National
Superconducting Cyclotron Laboratory (NSCL) at Michigan State
University and advances in experimental techniques \cite{TB-N07}.
Indeed, recent measurements
at the NSCL~\cite{OT-PRC07,TB-N07,PFM-BAPS08,OT-PRC09} have demonstrated that
the fragmentation of $^{48}$Ca and $^{76}$Ge beams can be used to
produce new isotopes in the proximity of the neutron dripline.
Continuing this work, we report here the next step with a newly developed $^{82}$Se beam towards the
fundamental goal of defining the absolute mass limit for chemical
elements in the region of calcium.
In the present measurement, four neutron-rich
isotopes with \protect{$42\le N\le 47$} were identified for the
first time (see Fig.\ref{chart}), one event was registered consistent with $^{70}$Cr$_{46}$, and another one with
$^{75}$Fe$_{49}$.

%WWWWWWWWWWWWWWWWWWWWWWWWWWWWWWWWWWWWWWWWWWWWWWWWWWWWWWWWWWWWWWWWWWWWWWWWWWWWWW

%WWWWWWWWWWWWWWWWWWWWWWWWWWWWWWWWWWWWWWWWWWWWWWWWWWWWWWWWWWWWWWWWWWWWWWWWWWWWWW

\subsection{Evidence for  global structure changes}

One of the first indications of significant changes in the structure of neutron rich nuclei was the discovery of enhanced nuclear binding of heavy sodium isotopes~ \cite{CT-PRC75}. This is now understood to result from significant contributions of {\it fp} shell intruder orbitals to the ground-state configurations of these isotopes~ \cite{XC-NPA75,EKW-PRC90}. Low-lying 2$^+$ states and quadrupole collectivity have been reported in neutron-rich even-even Ne and Mg isotopes around $N=20$, see for example Refs.~\cite{DGM-NPA84,MO-PLB95,YO-PLB01,YA-PLB03,CH-PRC05}.
This region around $^{31}$Na, where the neutron {\it fp} shell contributes significantly to the ground-state structure, is now known as the ``Island of Inversion''. Similarly, there is mounting evidence for an onset of deformation around neutron number $N=40$ in Fe and Cr nuclei. In even-even Fe and Cr nuclei, for example, this evidence is based on the energies of low-lying states~\cite{HA-PRL99,SO-EPJA03,AG-PRC08,AO-PRL09,GA-PRC10}, transition strengths~\cite{RO-PRL11}, deformation length ~\cite{AO-PRL09}, and higher-spin level schemes~\cite{ZH-PRC06}. Neutron $g_{9/2}$ and $d_{5/2}$ configurations from above the $N=40$ shell gap are proposed to descend and dominate the low-lying configurations similar to those in the $N=20$ Island of Inversion~\cite{BAB-PPNP01,LE-PRC10}.

Recently, it was shown that the beta-decay half-lives of the neutron-rich Ca isotopes~\cite{PFM-PRC08} compare favorably with the results of shell-model calculations performed in the full {\it pf} model space using the GXPF1 effective interaction~\cite{HO-EPJA05}. The systematic trend of these half-lives is consistent with the presence of a subshell gap at $N = 32$ as predicted by this interaction and confirmed by a variety of experiments. This interaction also predicts an increase of the excitation energy of the first excited  state $E_x(2_1^+)$ at $^{54}$Ca relative  to that obtained with the KB3G~\cite{KB3G-01}
interaction, suggesting the appearance of the $N = 34$ shell gap in Ca isotopes. Both interactions predict similar structure for light stable nuclei, but  give rather different predictions for several cases of neutron-rich nuclei.

Recent measurements at RIKEN of the $E_x(2_1^+)$  in  $^{54}$Ca~\cite{RIKEN-54Ca} found that the experimental value is 0.5 MeV smaller than the GXPF1B prediction, where the GXPF1B~\cite{GXPF1B} Hamiltonian was created from the GXPF1A Hamiltonian
 by changing five $T = 1$ matrix elements and the single-particle energies which involved $1p_{1/2}$.   A similar trend had already been pointed out by Mantica \emph{et al.}~\cite{PFM-PRC08} where they deduce that the effective energy gap between the adjacent neutron single-particle orbitals $f_{5/2}$ and $p_{1/2}$ is overestimated by the GXPF1 and GXPF1A effective interactions.  Based on this, the GXPF1B interaction has been modified to correct this 0.5~MeV shift and is referred to as GXPF1B5 here.

Other evidence supports the modified form of the  GXPF1B interaction. The original GXPF1B interaction predicts a 1n-unbound $^{56}$K ($S_{1n} = -0.03$~MeV); however, this isotope was shown to be bound by observation  in our previous experiment with the $^{76}$Ge beam.  The shift in the interaction makes the isotopes with valence neutrons in the $f_{5/2}$ orbital around $Z=20$ more bound, such that the modified interaction GXP1FB5 predicts a bound $^{56}$K with $S_{1n} = 0.41$~MeV.

In our previous cross section measurements in the region around $^{62}$Ti ($^{76}$Ge primary beam)~\cite{OT-PRL09}
we observed  a systematic  variation of the production cross sections that might point to nuclear structure effects,
such as an onset of collectivity, that are not included in global mass models that were  used to construct the basis of the systematics.
The present work,
since it is based on isotope production from  a different primary beam,
 covering the same region of the nuclear chart,
provides an independent check of this interpretation.

%WWWWWWWWWWWWWWWWWWWWWWWWWWWWWWWWWWWWWWWWWWWWWWWWWWWWWWWWWWWWWWWWWWWWWWWWWWWWWW

\begin{table*}
\caption{ Experimental settings}\label{Tab_runs}
\begin{tabular}{|c|c|ccccc|rl|rl|c|c|c|c|c|c|}
\hline
       Data & {\small{Fragment}}  &     \multicolumn{ 5}{|c|}{Magnetic rigidity, $B\rho (Tm)$} & \multicolumn{ 2}{|c|}{Target} &  \multicolumn{
       2}{|c|}{Stripper}
& Wedge  &   $\Delta p/p$ & Time &  Beam  &  Goal \\

           set &  {\footnotesize{of interest}} &     $D_1D_2$ &     $D_3D_4$ &      $D_5D_6D_7$ &     $D_8D_9$ &      $D_{10}D_{11}$ &
            \multicolumn{ 2}{|c|}{\footnotesize{$mg/cm^2$}} &  \multicolumn{ 2}{|c|} {\footnotesize{$mg/cm^2$}} &     {\footnotesize{$mg/cm^2$}} &       (\%) &
                  {\footnotesize{$hours$}} &         particles &            \\
\hline

         1 &        $^{67}$Fe &     ~4.3209 &     ~4.3209 &     ~4.3065 &     ~4.2919  &       ~4.2867 &         ~Be &        9.7 &    &     -   &      -      &        0.1 &       1.13 &        {\footnotesize{3.76e12}} & \multicolumn{ 1}{|c|}{} \\

         2 &        &      &      &      &     &        &         Be &       68 &     &   -    &     -       &        0.1 &       1.01 &         {\footnotesize{3.06e12}} & \multicolumn{ 1}{|c|}{} \\

         3 &        &      &      &      &     &        &         Be &        138 &     &    -   &      -      &        0.1 &       0.69 &         {\footnotesize{4.32e13}} & \multicolumn{ 1}{|c|}{\footnotesize{momentum}} \\

         4 &        &      &      &      &     &        &         Be &        230 &      &    -  &      -      &        0.1 &       1.17 &         {\footnotesize{2.00e14}} & \multicolumn{ 1}{|c|}{\footnotesize{distribution}} \\

         5 &        &      &      &      &     &        &         Be &        314 &       &   -  &      -      &        0.2 &       1.03 &         {\footnotesize{1.05e14}} & \multicolumn{ 1}{|c|}{\footnotesize{study}} \\

         6 &        &      &      &      &     &        &         Be &        413 &       &   -  &      -      &        0.2 &       1.45 &         {\footnotesize{2.60e14}} & \multicolumn{ 1}{|c|}{} \\

         7 &        &      &      &      &     &        &         Be &        513 &       &   -  &      -      &        0.2 &       1.60 &         {\footnotesize{4.13e14}} & \multicolumn{ 1}{|c|}{} \\

\hline

         8 &        $^{67}$Fe &     4.3412 &     4.3209 &     4.3065 &     4.2919 &        4.2867 &         Be &        190 &        &  -  &         20 &          0.2 &    1.99 &     {\footnotesize{1.14e15}} & \multicolumn{ 1}{|c|}{\footnotesize{isomer}}\\

         9 &        $^{78}$Zn &     4.3505   &   4.3267  &   4.3099   &    4.2928 &        4.2867 &         Be &        190 &        &  -  &         20 &          0.2 &    2.00 &       {\footnotesize{1.57e15}} & \multicolumn{ 1}{|c|}{\footnotesize{production}} \\

\hline

         10 &        $^{74}$Fe &     4.3538 & 	4.3289 & 	4.3111 & 	4.2931 & 	4.2867 &         Be &        557 &       &  -   &        20 &          5 &      37.4 &        {\footnotesize{3.22e16}} & \multicolumn{ 1}{|c|}{} \\

         11 &        $^{75}$Fe &     4.3560 & 	4.3301 & 	4.3118 & 	4.2933 & 	4.2867 &          W &        750 &    Be & 17.3 &        20 &          5 &      3.86 &        {\footnotesize{3.66e15}} & \multicolumn{ 1}{|c|}{\footnotesize{production}} \\

         12 &        $^{68}$V  &     4.3515 & 	4.3274 & 	4.3103 & 	4.2929 & 	4.2867 &         Be &        695 &       & -    &        20 &          5 &      42.6 &      {\footnotesize{3.77e16}} & \multicolumn{ 1}{|c|}{\footnotesize{of new }} \\

         13 &        $^{60}$Ca &     4.3451 & 	4.3233 & 	4.3079 & 	4.2922 & 	4.2867 &         Be &        849 &       &  -   &        20 &          5 &      16.1 &        {\footnotesize{1.49e16}} & \multicolumn{ 1}{|c|}{\footnotesize{isotopes}} \\

         14 &        $^{60}$Ca &     4.3451 & 	4.3233 & 	4.3079 & 	4.2922 & 	4.2867 &         Be &        695 &       &  -   &        20 &          5 &      14.8 &        {\footnotesize{1.18e16}} & \multicolumn{ 1}{|c|}{} \\

\hline
        15 &         $^{45}$Ca &     3.6331 & 	3.6177 & 	3.6055 & 	3.5932 & 	3.5888 &         Be &        190 &       &  -   &        20 &          0.1 &     0.92 &      {\footnotesize{2.86e11}} & \multicolumn{ 1}{|c|}{\footnotesize{stable}} \\

        16 &         $^{48}$Ca &     3.6396 & 	3.6219 & 	3.6080 & 	3.5939 & 	3.5888 &         Be &        190 &       &  -   &        20 &          0.1 &     1.55 &      {\footnotesize{2.50e11}} & \multicolumn{ 1}{|c|}{\footnotesize{Ca isotopes}} \\
\hline
\end{tabular}
\end{table*}

%WWWWWWWWWWWWWWWWWWWWWWWWWWWWWWWWWWWWWWWWWWWWWWWWWWWWWWWWWWWWWWWWWWWWWWWWWWWWWW

\section{Experiment\label{secExpt}}
\subsection{Setup\label{secSetup}}

A newly developed 139 MeV/u $^{82}$Se beam with an intensity of 35~pnA, accelerated by the coupled cyclotrons at
the NSCL, was fragmented in a series of beryllium targets and a
tungsten target, each placed at the object position of the A1900
fragment separator~\cite{DJM-NIMA03}. In this work we used an identical configuration to our previous experiment with a $^{76}$Ge beam~\cite{OT-PRC09},
where the combination of the A1900 fragment separator with
the S800 analysis beam line~\cite{DB-NIMB03} formed a two-stage
separator system, that allowed a high
degree of rejection of unwanted reaction products. At the end of the S800 analysis beam line, the
particles of interest were stopped in a telescope of eight silicon
PIN diodes (50$\times$50~mm$^2$ each) with a total thickness of 8.0~mm.  A 50~mm thick plastic scintillator
positioned behind the Si-telescope served as a veto detector against
reactions in the Si-telescope and provided a measurement of the
residual energy of lighter ions that were not stopped in the
Si-telescope. A position sensitive parallel plate avalanche counter
(PPAC) was located in front of the Si-telescope.
All experimental details and a sketch of the experimental setup can be found in Ref.~\cite{OT-PRC09}.
In this paper, we describe the details of our  experimental approach and discuss the
results.

\subsection{Experimental runs\label{secPlanning}}

The present experiment consisted of four  segments that are summarized
in Table \ref{Tab_runs}. Except for the last segment, the present experimental program
is similar to the previous  $^{76}$Ge experiment~\cite{OT-PRC09}.
During all runs, the magnetic rigidity of
the last two dipoles of the analysis line was kept constant at a value of 4.2867~Tm while the
production target thickness was varied to map the fragment momentum
distributions. This approach  greatly simplifies the particle
identification during the scans of the parallel momentum
distributions.

The   momentum acceptance of the A1900 fragment separator was
restricted to $\Delta  p/p = 0.1\%$ (first four runs with thin targets), and to $\Delta  p/p = 0.2\%$
 (other  targets) for the measurement of differential momentum
distributions in the first part of the
experiment. The use of different beryllium target thicknesses
(9.7, 68, 138, 230, 314, 413, 513~mg/cm$^2$) allowed coverage of the fragment momentum distributions necessary to extract
production cross sections and also resulted in more isotopes in the
particle identification spectrum.

 For the second part of the experiment, a Kapton wedge
with a thickness of 20.0 mg/cm$^2$ was used at the dispersive image of
the A1900  with  a 10~mm aperture in
the focal plane to reject less exotic fragments while the separator was set for $^{67}$Fe and $^{78}$Zn ions. The goal
of this setting was to confirm the particle identification by isomer
tagging as described in
Ref.~\cite{RG-PLB95} with $^{67m}$Fe~($E_{\gamma}=367$~keV, $T_{1/2}=43$~$\mu$s)
and $^{78m}$Zn~($E_{\gamma}=730, 890, 908$~keV, $T_{1/2}=0.32$~$\mu$s).

In the third part of the experiment, dedicated to the  search  for
 new isotopes, five settings were  used to cover the most
neutron-rich isotopes  with \protect{$20\le Z\le 27$}, as it was
impossible to find a single target thickness and magnetic rigidity
to observe all of the fragments of interest. Each setting was
characterized by a fragment for which the separator was optimized. A
search for the most exotic nuclei in each setting was carried out
with  Be and W targets. The settings were centered on $^{60}$Ca,
$^{68}$V and $^{74,75}$Fe respectively, based on
\liseppsh~\cite{OT-NIMB08} calculations using the parameterizations
of the momentum distributions obtained in the first part of the
experiment (see Section~\ref{secMomentum}). The momentum acceptance
of the A1900 was set to the maximum of $\Delta p/p = 5.0\%$ for
these production runs. It should be noted that the momentum acceptance of the S800 beamline  is about 4\%
according to \lisepp Monte Carlo simulations using a new extended configuration with 5$^{\text{th}}$ order optics.
This calculated acceptance has been used for the cross section analysis using the method described below.

The fourth part of the experiment was devoted to two short runs measuring the yields of more stable isotopes
by centering on $^{45,48}$Ca.

\begin{figure}
\centering
\includegraphics[width=0.95\columnwidth]{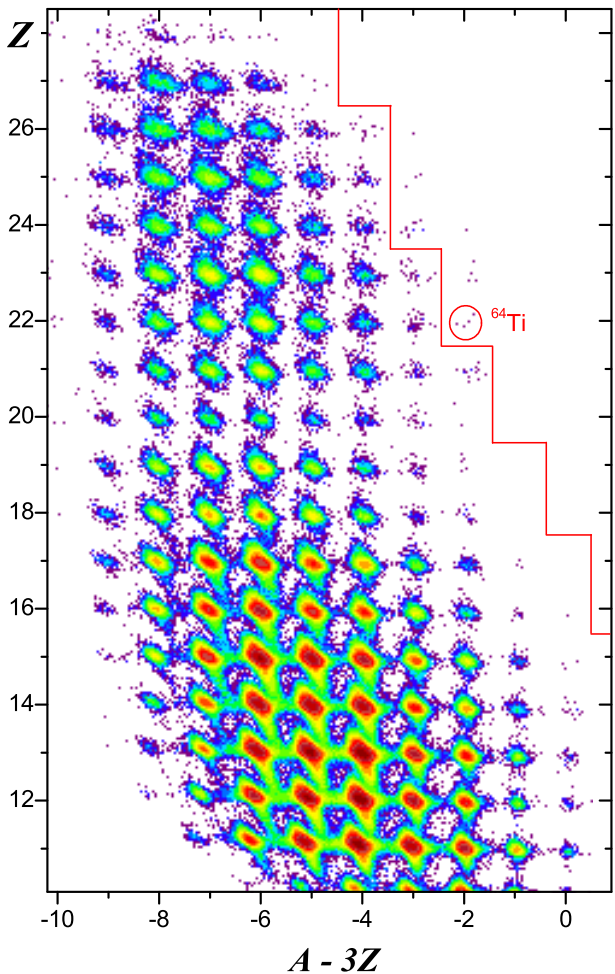}%
\caption{(Color online) Particle identification plot showing the
measured atomic number, $Z$, versus the calculated function $A-3Z$
for the nuclei observed in production runs of this work. See text for details. The limit of previously observed
nuclei is shown by the solid red line, and the location of
$^{64}$Ti is marked. \label{Fig_pid}}
\end{figure}

%WWWWWWWWWWWWWWWWWWWWWWWWWWWWWWWWWWWWWWWWWWWWWWWWWWWWWWWWWWWWWWWWWWWWWWWWWWWWWW
\section{Analysis of experimental data \label{secAnalysis}}

%WWWWWWWWWWWWWWWWWWWWWWWWWWWWWWWWWWWWWWWWWWWWWWWWWWWWWWWWWWWWWWWWWWWWWWWWWWWWWW

\begin{SCfigure*}
\includegraphics[width=0.82\textwidth]{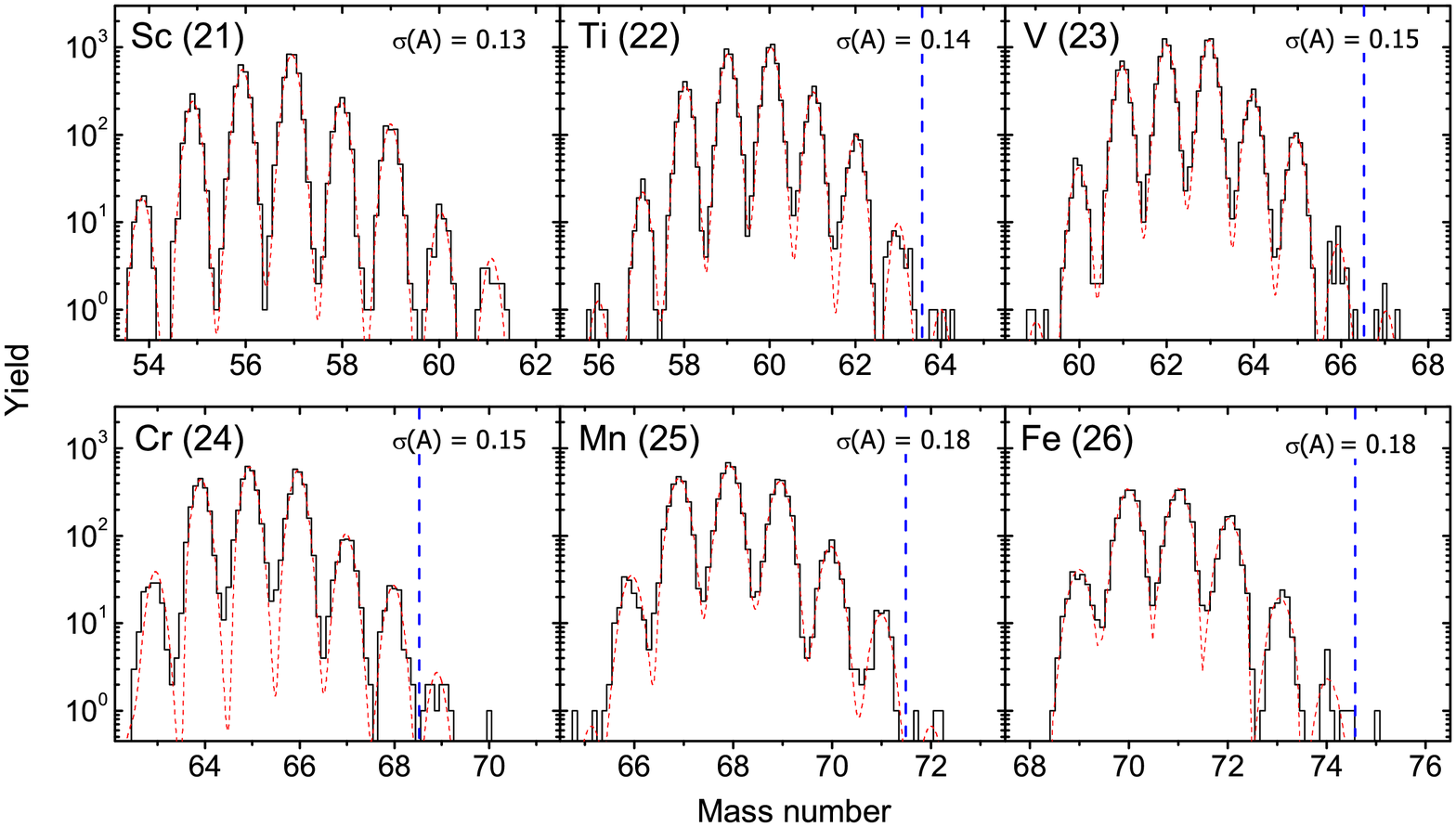}%
\caption{(Color online) Mass spectra of the elements \protect{$21\le Z\le 26$}
  that were stopped in the Si-telescope
during the production runs. The limits of previously
observed nuclei are shown by the vertical dashed lines.
Standard deviations produced  with the Gaussian
function at constant width (dashed curves) are given  in the
panel figures for each element. \\ \\
\label{Fig_Mass}}
\end{SCfigure*}

%WWWWWWWWWWWWWWWWWWWWWWWWWWWWWWWWWWWWWWWWWWWWWWWWWWWWWWWWWWWWWWWWWWWWWWWWWWWWWW
%\subsection{Search for new isotopes\label{secNewIsotopes}}

The result of our approach of keeping the last dipoles magnetic rigidities
 constant while varying the target thickness
--- as was done in the previous experiment --- can be seen in Fig.~\ref{Fig_pid}, which shows the total
distribution of fully-stripped reaction products observed in the production runs
of this work. The range of fragments is shown as a function of the measured
atomic number, $Z$,  versus the quantity $A-3Z$ deduced from measured values, where $A$ is the mass number.
The identification of the individual isotopes in Fig.~\ref{Fig_pid}
was confirmed via isomer tagging using the known isomeric decays in
$^{67}$Fe and $^{78}$Zn. The
standard deviations of ionic charge ($q$) and elemental ($Z$) spectra  were found to be similar to those in the previous experiment,
therefore the probabilities of one event being
misidentified as a neighboring charge state or element as before~\cite{OT-PRC09}.
The details of the calculation of the particle identification are given
in the appendix to the previous work~\cite{OT-PRC09}.

The mass spectra for the isotopic chains from scandium to iron
measured during the production runs are shown in
Fig.~\ref{Fig_Mass}. Only nuclei that stopped in the Si telescope are included in this analysis. The observed fragments include
several new isotopes that are the most neutron-rich nuclides yet observed
of elements \protect{{$22\le Z\le 25$} ($^{64}$Ti, $^{67}$V, $^{69}$Cr,
$^{72}$Mn)}. One event was found to be consistent with $^{70}$Cr, and another one with
$^{75}$Fe.  The new neutron-rich nuclei
observed in this work lie to the right of the solid
line in Fig.~\ref{Fig_pid} and to the right of the vertical dashed
lines in Fig.~\ref{Fig_Mass}.

%WWWWWWWWWWWWWWWWWWWWWWWWWWWWWWWWWWWWWWWWWWWWWWWWWWWWWWWWWWWWWWWWWWWWWWWWWWWWWW
%WWWWWWWWWWWWWWWWWWWWWWWWWWWWWWWWWWWWWWWWWWWWWWWWWWWWWWWWWWWWWWWWWWWWWWWWWWWWWW
\section{Results and Discussion\label{secRes}}

\subsection{Parallel momentum distributions\label{secMomentum}}

The prediction of the momentum distributions of
residues is important when searching for new isotopes  in order to set the
fragment separator at the maximum production rate. Also, the
accurate prediction of the momentum distributions allows for a precise estimate
of the transmission and efficient rejection of strong contaminants.
In this experiment the ``target scanning" approach~\cite{OT-NIMA09}, developed in the previous experiment,
was used to obtain parameters for the neutron-rich isotope momentum distribution models such as ~\cite{AG-PLB74,DJM-PRC89}.
This method is particularly well suited to survey
neutron-rich nuclei since the less exotic nuclei are produced with
the highest yields and their momentum distributions can be readily measured
with thin targets.

The data analysis of this approach has been improved, and a detailed description is in preparation~\cite{OT-NIMA12}.
Important improvements include: first,  the most probable velocity for a fragment is not
that at the center of the target when the yield is sharply rising  or falling with momentum,
and second, asymmetric Gaussian distributions have been used with asymmetry coefficients
 taken from the convolution model~\cite{OT-NPA04} implemented in the \lisepp code~\cite{OT-NIMB08}. Note that, at the bombarding energy used in these experiments,
 the shape of the fragment momentum distribution is asymmetric with a
low-energy exponential tail stemming from dissipative
processes~\cite{OT-NPA04}.

\begin{figure}[b]
\includegraphics[width=1.0\columnwidth]{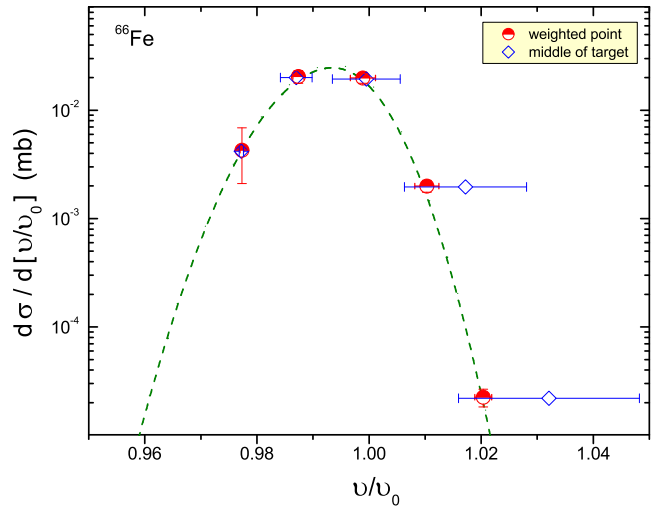}%
\caption{(Color online) Differential cross sections of $^{66}$Fe
 fragments obtained by varying the
target thickness with the analysis line set at one fixed magnetic rigidity. The dashed line represents
the fitted asymmetric Gaussian function. Blue horizontal errors with diamonds in center correspond to the
velocity difference caused by production at the beginning or the end
of target, whereas the red circles show the position of the most probable velocity based on the momentum distribution parametrization.\label{Fig_Velocity}}
\end{figure}

\begin{figure}[b]
\includegraphics[width=1.0\columnwidth]{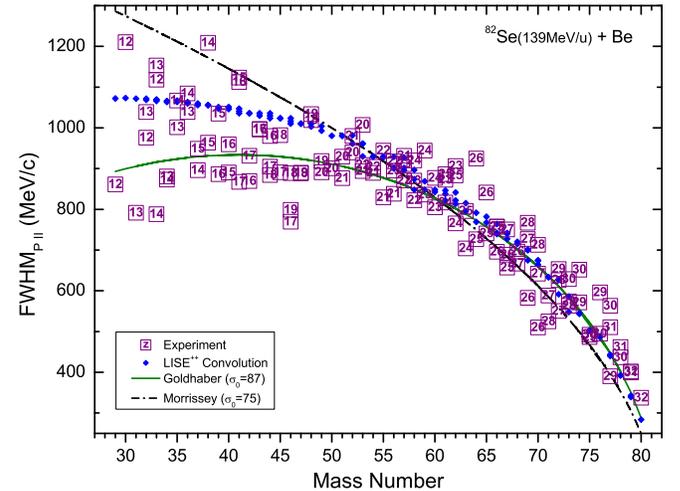}%
\caption{(Color online) Widths of parallel momentum component  as a function of  mass number of fragments produced in the reaction $^{82}$Se beams with  beryllium targets. Small diamonds denote calculations by the convolution model~\cite{OT-NPA04} with default settings for separation energy~($E_s$) option \#1  in \liseppsh. Solid green and dot-dashed black lines represent the best fit to the data for the Goldhaber~\cite{AG-PLB74} and Morrissey~\cite{DJM-PRC89} models, respectively.\label{Fig_FWHM}}
\end{figure}

Seven targets were used to measure the momentum distributions (see
Table~\ref{Tab_runs}).  The yield of one example fragment,
$^{66}$Fe, is shown in Fig.~\ref{Fig_Velocity} as a
function of the ratio of  fragment and beam velocities.
This figure illustrates the impact of  the new data analysis where the most probable values taken are shown by the circles and the average values by the diamonds.
Momentum distributions for 126 isotopes were derived (indicated by the colored boxes
in Fig.~\ref{chart}) and integrated to deduce the production cross sections.

A survey of all of the fitted results showed that fragments in the
heavy mass region were produced similar to our previous measurements~\cite{OT-NIMA09} with significantly higher velocities
than the momentum distribution models~\cite{DJM-PRC89,BOR-ZPA83} predict.
The difference is most likely due to the fact that the
models were developed for fragments close to stability, where the energy
required to remove each nucleon was set to 8~MeV, while the actual nucleon binding energy for the neutron rich isotopes under investigation is lower.
An analysis with asymmetric distributions to reproduce the mean velocity of fragments
has shown that the neutron-rich separation energy parameter in the model ~\cite{DJM-PRC89} for the nuclei
observed in the present work in the region $A_P/2 \le A_F \le A_P$
can be represented by a linear decrease with the number of removed nucleons:
\begin{equation} \label{Eq_Velocity}
E_S  = 8 - 9.2 \Delta A / A_P
\end{equation}
where $\Delta A = A_P - A_F$, $A_P$ is the projectile mass number,
and $A_F$ is the fragment mass number.
In the fourth part of the experiment, where stable isotopes were measured (see Table \ref{Tab_runs}),
no deviations from the  default parameters of the model for the velocities were observed.

The width obtained for each fragment's parallel momentum distribution is presented in Fig.~\ref{Fig_FWHM} for fragments produced from the interaction of $^{82}$Se~(139~MeV/u) with $^9$Be targets.  The predictions with
best fits of reduced widths   $\sigma_0(G)=87$~MeV/$c$ for the Goldhaber~\cite{AG-PLB74} and  $\sigma_0(M)=75$~MeV/$c$ for the Morrissey~\cite{DJM-PRC89} models are presented in this figure.

%------------------------------------------------------------------------------------------------------

%WWWWWWWWWWWWWWWWWWWWWWWWWWWWWWWWWWWWWWWWWWWWWWWWWWWWWWWWWWWWWWWWWWWWWWWWWWWWWW

\begin{figure*}
\includegraphics[width=1.0\textwidth]{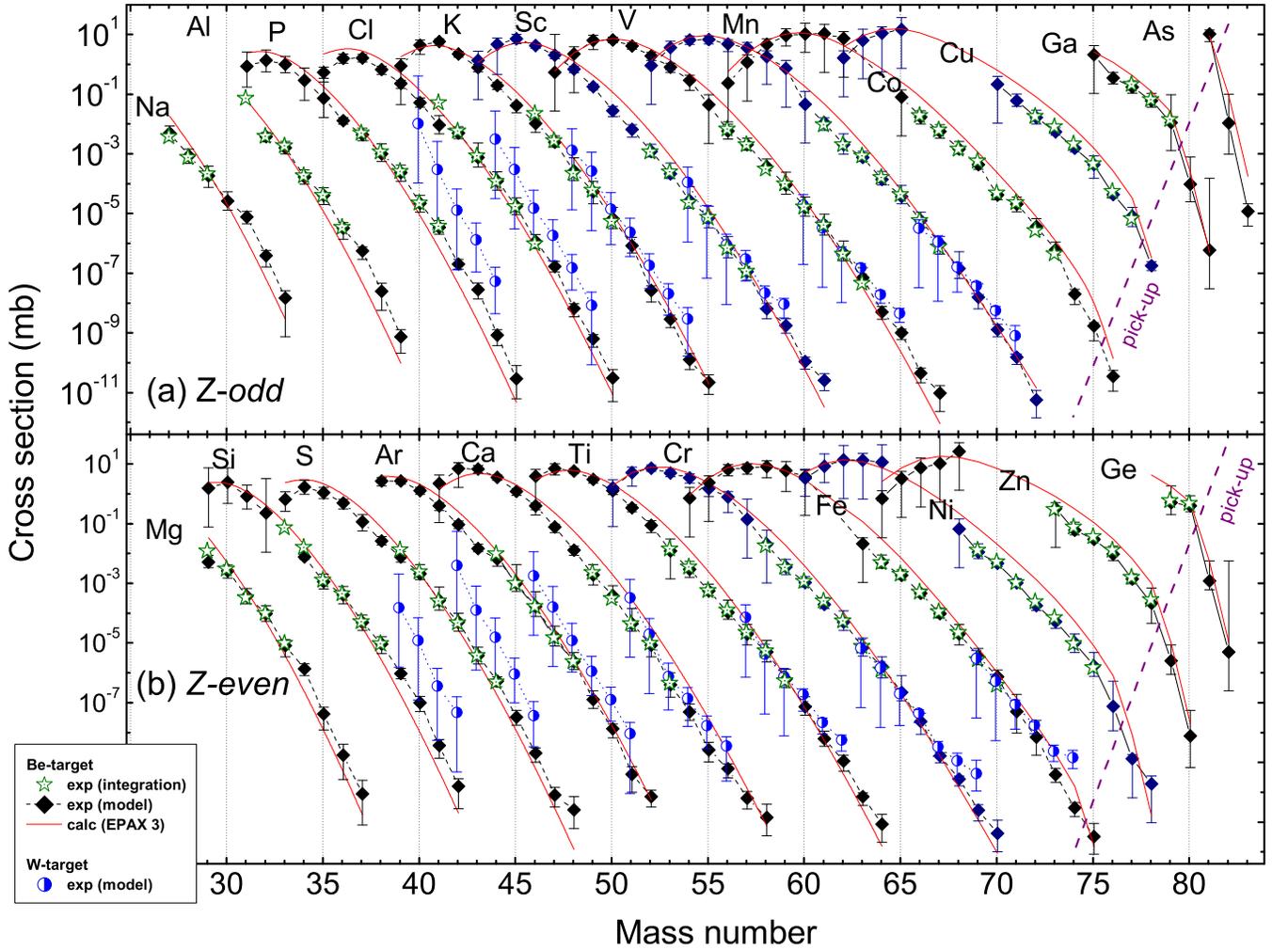}%
\caption{(Color online) Inclusive production cross sections for
fragments from the reaction of $^{82}$Se with beryllium and tungsten
targets shown as a function of mass number. The cross sections with
the beryllium targets derived by momentum distribution integration
are shown by stars, those normalized with \liseppsh\ transmission
calculations are indicated by solid diamonds The cross sections
obtained with the tungsten target were normalized with \liseppsh\
transmission calculations.  The
red solid lines show the predictions of the \epaxthree systematics~\cite{KS-EPAX3} for
beryllium (see text).
The two magenta dashed lines separate
nuclei that require neutron pickup in the production mechanism.
\label{Fig_CrossSection}}
\end{figure*}

%WWWWWWWWWWWWWWWWWWWWWWWWWWWWWWWWWWWWWWWWWWWWWWWWWWWWWWWWWWWWWWWWWWWWWWWWWWWWWW

\subsection{Production cross sections\label{secCS}}

The inclusive production cross sections for the observed fragments
were calculated by correcting the measured yields for the finite
momentum and angular acceptances of the separator system. A
total of 126 cross sections with beryllium targets were obtained
from the Gaussian fits to the longitudinal momentum
distributions; these nuclei are indicated by stars in
Fig.~\ref{Fig_CrossSection}. The cross sections for all of the
remaining fragments with incompletely measured longitudinal momentum
distributions were obtained with estimated transmission corrections as in our previous work~\cite{OT-PRC09}.
The parameters for the transmission corrections were assumed to be smoothly varying with $A$ and $Z$.

The cross sections obtained for all the fragments observed in
this experiment are shown in Fig.~\ref{Fig_CrossSection} along with
the predictions of the recent \epaxthree parameterization~\cite{KS-EPAX3}.  For those isotopes that relied on
transmission calculations, the weighted mean of all measured yields
was used to obtain the  cross section (shown by solid
diamonds in Fig.~\ref{Fig_CrossSection}). The uncertainties in these
cases include the statistical, the systematic and the transmission
correction uncertainties. For more details see ref.~\cite{OT-NIMA09}.
As can be seen in Fig.~\ref{Fig_CrossSection}, the
 cross sections are in good agreement with those produced
by integrating the measured longitudinal momentum distributions  in the cases  where there is an overlap.

It is important to note that the  predictions
of the  recent \epaxthree parameterization for reactions with beryllium, shown
by the solid lines in Fig.~\ref{Fig_CrossSection}, reproduces the
measured cross sections for isotopes much better than the previous \epaxtwo predictions~\cite{KS-PRC00}.

%WWWWWWWWWWWWWWWWWWWWWWWWWWWWWWWWWWWWWWWWWWWWWWWWWWWWWWWWWWWWWWWWWWWWWWWWWWWWWW

\begin{figure}[t]
\includegraphics[width=0.94\columnwidth]{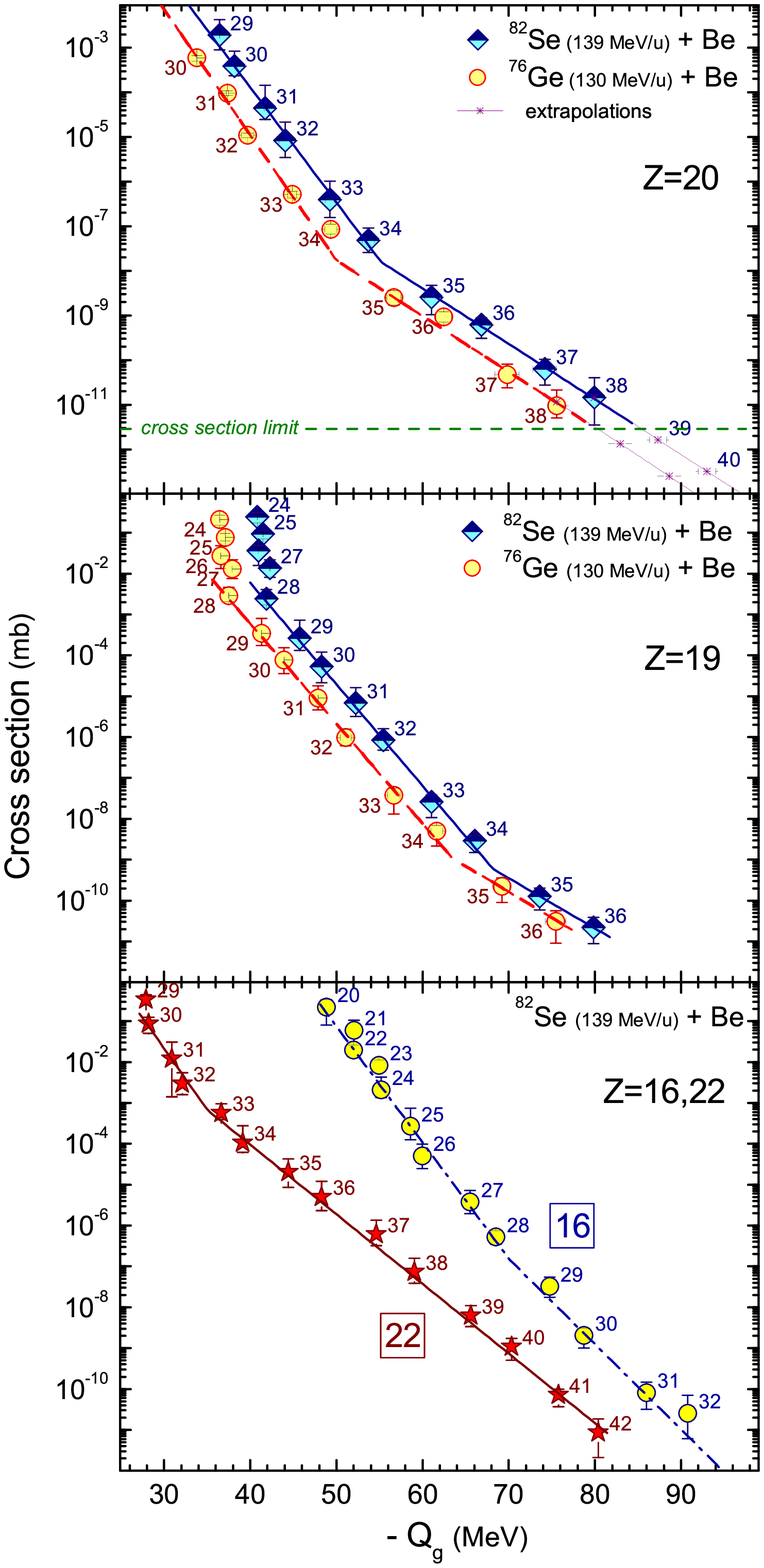}%
\caption{(Color online) Measured cross sections versus  -$Q_{\text{g}}$  for the production
of neutron-rich calcium (top plot) and potassium (middle plot) isotopes from reactions of $^{82}$Se and $^{76}$Ge with
a beryllium target. Production cross section of sulfur and titanium isotopes from the reaction of $^{82}$Se with a beryllium target are shown in bottom panel.
See text for explanation of $Q_g$ and the lines.  Neutron numbers of isotopes are shown in figure.
A horizontal dash green line in top plot shows the cross section limit reached in the experiment with the $^{82}$Se beam.
\label{Fig_Qg_all}}
\end{figure}

\begin{figure}[t]
\includegraphics[width=0.94\columnwidth]{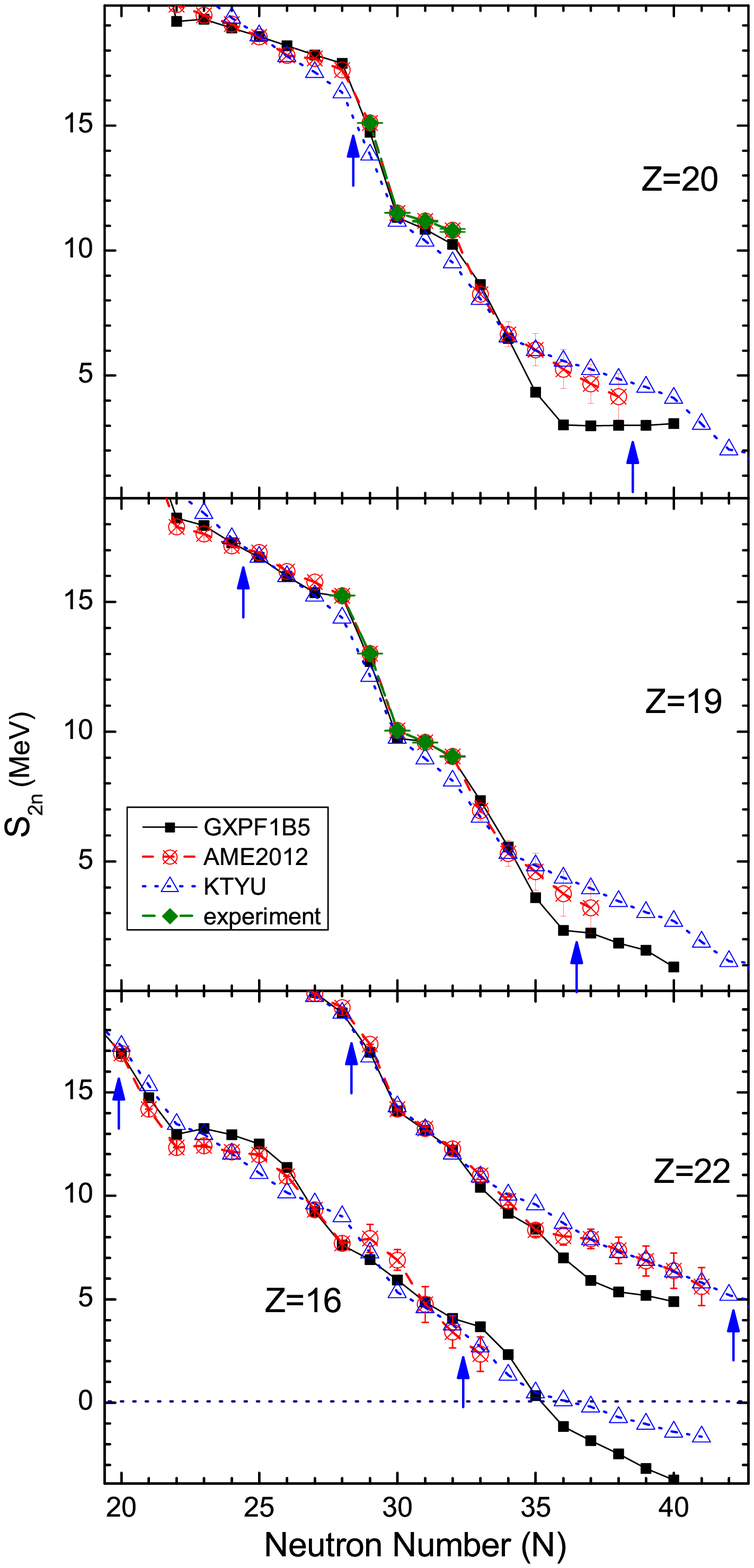}%
\caption{(Color online) The two-neutron separation energy $S_{2n}$ deduced
from mass values as function of neutron number for calcium (top), potassium (middle), sulfur and titanium isotopes (bottom).
Those from experimental mass values~\cite{GA-PRL12} are shown by diamonds and from  AME2012~\cite{AME2012} by crossed circles.
Results based on the full $pf$-shell phenomenological
GXPF1B5~\cite{GXPF1B} interaction and the KTYU mass
model~\cite{KTUY-PTP05} are shown by solid squares and hollow triangles,
respectively. Arrows show regions of isotopes whose measured cross section are shown in Fig.~\ref{Fig_Qg_all}.
\label{Fig_Sn_All_models}}
\end{figure}

%WWWWWWWWWWWWWWWWWWWWWWWWWWWWWWWWWWWWWWWWWWWWWWWWWWWWWWWWWWWWWWWWWWWWWWWWWWWWWW

\subsection{Q$_g$ systematics\label{secQg}}

The production cross sections for the most neutron-rich projectile
fragments have been  previously shown to have an exponential
dependence on $Q_{\text{g}}$, where $Q_{\text{g}}$ is defined as the difference in mass-excess
between the beam particle and the observed fragment~\cite{OT-PRC07,OT-PRL09}.
To test this behavior, the cross sections for each isotopic chain were fitted
with the simple expression:
\begin{equation}\label{Eq_Qg}
 \sigma(Z,A) = k(Z)\exp{(Q_{\text{g}}(Z,A))/T(Z))},
\end{equation}
where $T$ represents an inverse slope parameter, and $k$ is a normalization.

\begin{figure}[t]
\includegraphics[width=1.0\columnwidth]{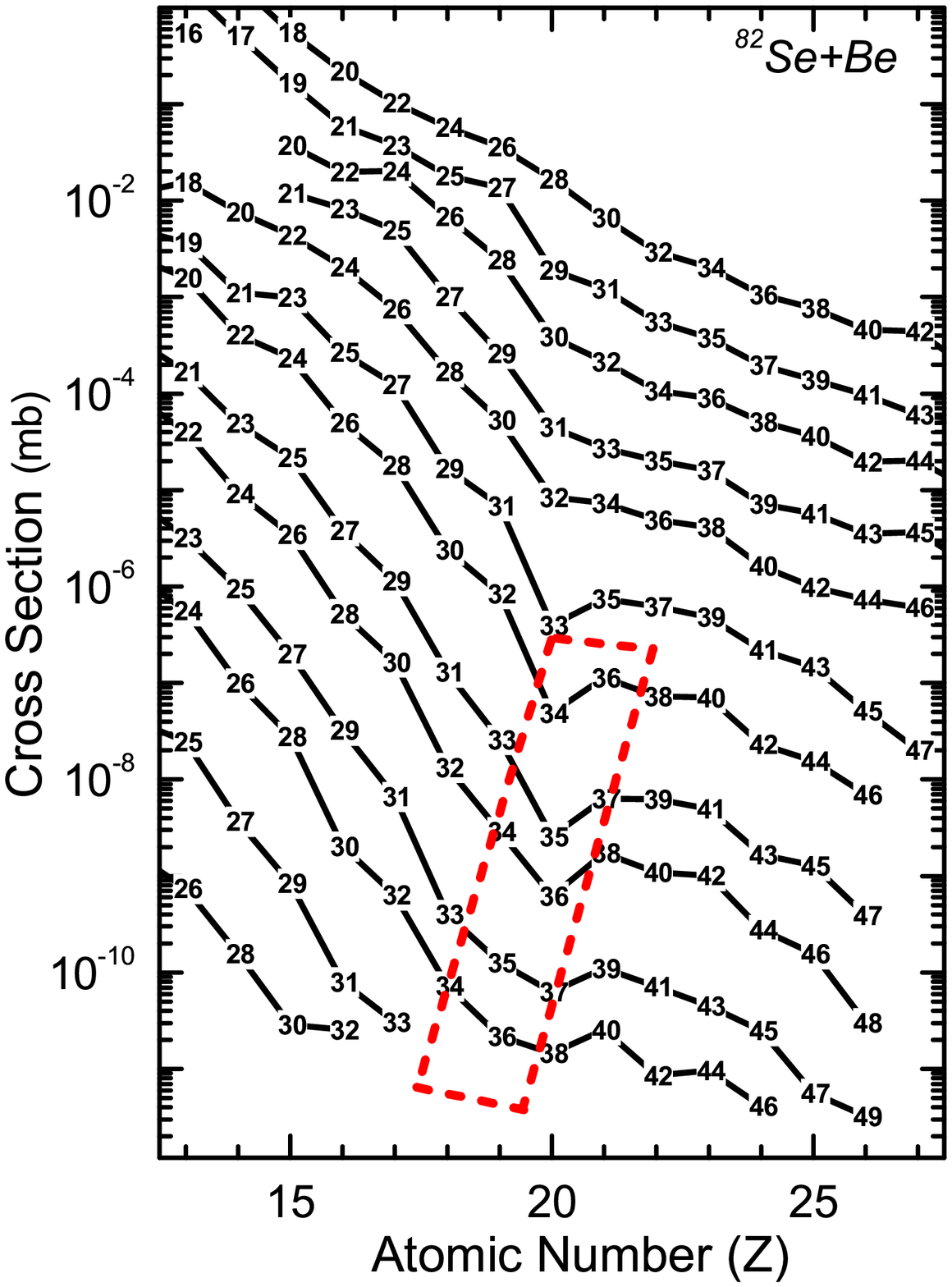}%
\caption{(Color online) Production cross section versus atomic
number ($Z$) for fragments from reaction of $^{82}$Se with beryllium targets.
Lines are connected according to constant $N-2Z$, while labels represent the
neutron number.
Reactions resulting in neutron pick-up are omitted.
The red dashed quadrangle is explained in the text. \\
\label{Fig_triton_plot}}
\end{figure}

\begin{figure}[t]
\includegraphics[width=1.0\columnwidth]{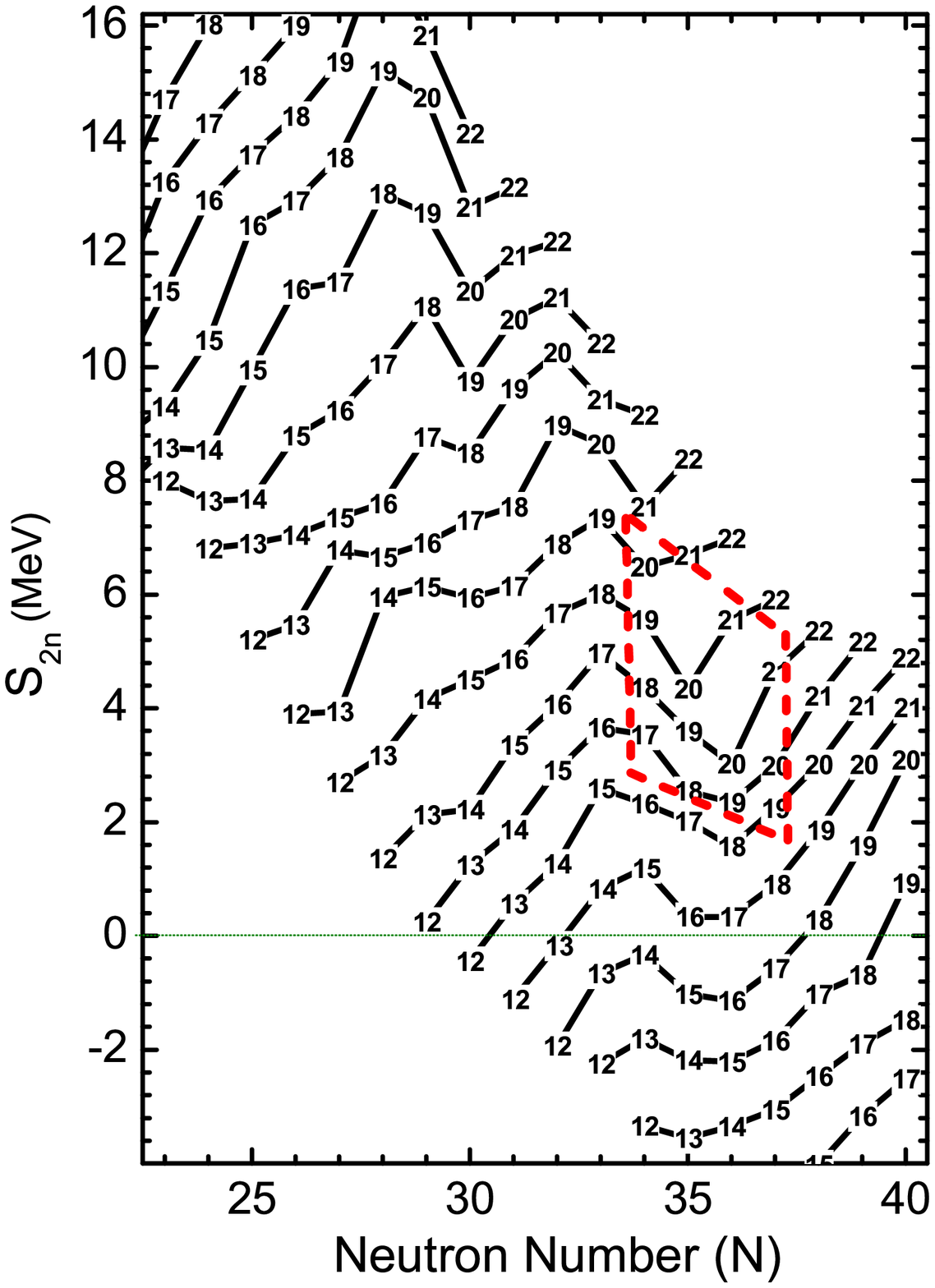}%
\caption{(Color online) Two-neutron separation energy $S_{2n}$
versus neutron number ($N$) for elements $12 \le Z \le 22$. Values are calculated using results from the
GXPF1B5~\cite{GXPF1B} model.
Labels in the lines show atomic numbers of nuclei.
The red dashed quadrangle is explained in the text.\\
\label{Fig_s2n_Zconnect}}
\end{figure}

Most of the data from the reactions of $^{82}$Se on Be targets in this experiment
could be fitted by two inverse slopes with a floating connection point.
The trends of the general increase in
 $T$ for all of the heavy isotopes of elements $Z=$~19, 20, and 21 observed with a $^{76}$Ge beam
 in our previous experiment is reproduced here with the $^{82}$Se beam.
The cases of $Z=$~16, 19, 20, and 22 are illustrated in Fig.~\ref{Fig_Qg_all} showing the
measured cross sections versus $Q_{\text{g}}$ calculated using the masses
deduced from the shell model with the GXPF1B5 interaction~\cite{GXPF1B}.
As in the previous experiment,  the heaviest isotopes of elements in the vicinity of $Z=$~20 appear to
deviate  from an exponential dependence.
The change of slope is most dramatic at $Z=20$.
At $Z=22$, the trends show  little change in slope for heaviest isotopes.
As was pointed out in our previous study~\cite{OT-PRL09}, a possible
explanation for the exponential slope reduction at larger masses for these elements
is a change in binding energy relative to predictions.
It should be noted that the systematic variation of the production cross sections of the calcium isotopes
as a function of $Q_{\text{g}}$ was checked in our previous work with several other
well-known mass models and essentially the same behavior
was observed (for details, see Fig.10 in~\cite{OT-PRC09}).

%=======================================================================================================================
%\begin{SCfigure*}
\begin{figure*}
\includegraphics[width=0.95\textwidth]{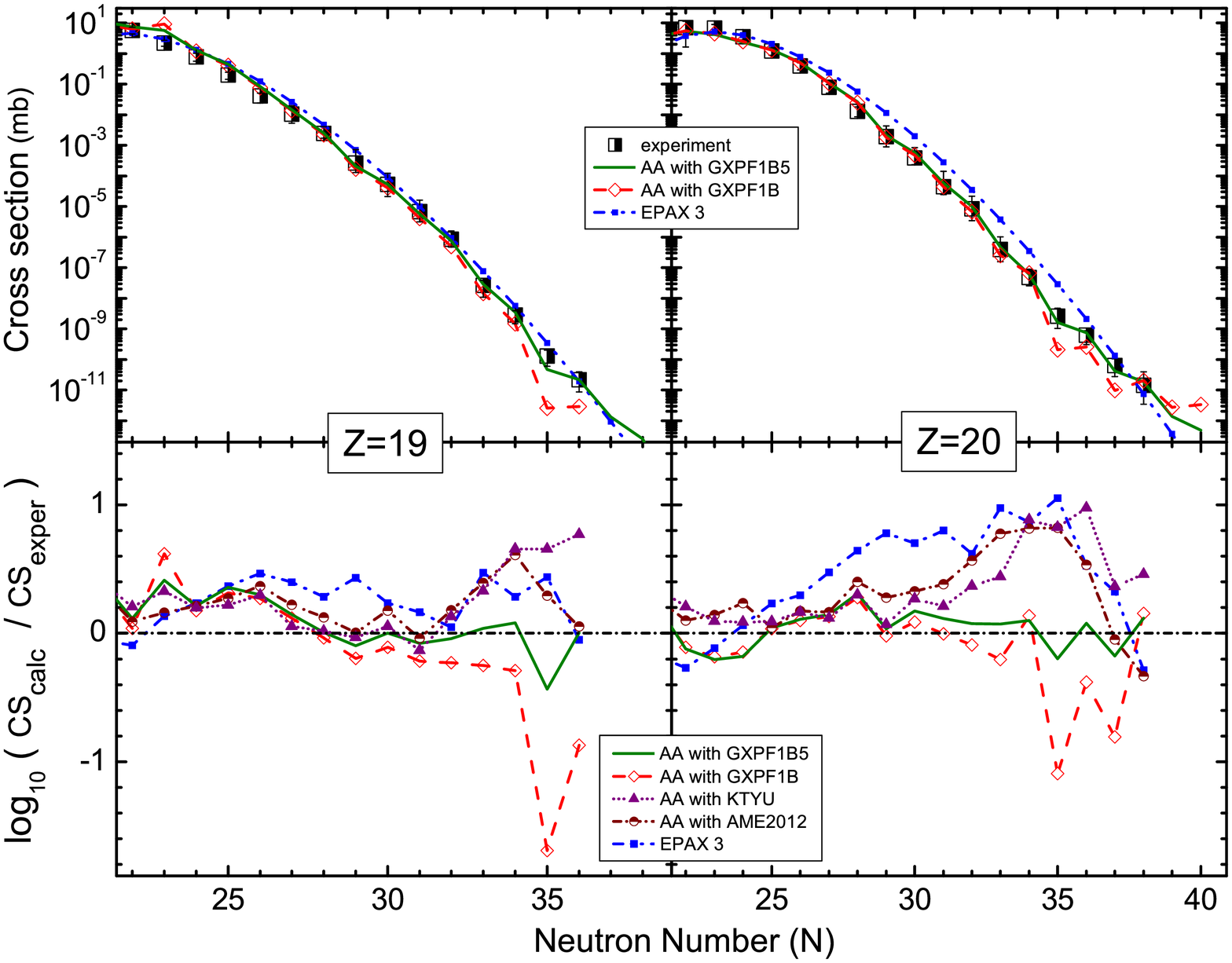}
\caption{(Color online) Top: The cross section versus neutron number for the
production of neutron-rich potassium (left) and calcium (right) isotopes from
the reaction of $^{82}$Se on beryllium targets.  The measured values are
shown by semi-solid rectangles.
The results from the Abrasion-Ablation calculations  using the masses
obtained via both the GXPF1B and GXPF1B5 interactions~\cite{GXPF1B} are shown by open diamonds with dashed lines and
solid lines correspondingly.
For comparison, the values obtained from the
\epaxthree systematics~\cite{KS-EPAX3} (small solid rectangles with dashed-dot-dot lines) are also included.\\
Bottom: Logarithm ratios of calculated and experimental cross sections shown  for clearer presentation differences between experimental and calculated values.
This presentation includes results from
the KTYU model~\cite{KTUY-PTP05} (solid triangles with dot lines) and the AME2012 table~\cite{AME2012} (semi-solid circles with dash-dot lines).
\label{Fig_CS_Z19_20}}
\end{figure*}
%=======================================================================================================================

%\subsection{Shell-model with GXPF1B5\label{secSn}}

 A hint at the origin of the cross section systematics may be seen in the binding energy trends demonstrated in Fig.~\ref{Fig_Sn_All_models}, where the dependence of $S_{2n}$ is shown as a function neutron number in the neutron-rich region  for sulfur,  potassium, calcium and titanium based on masses from models and experimentally measured values. There is no abrupt behavior for potassium and calcium in the KTYU model~\cite{KTUY-PTP05} or the AME2012 mass table~\cite{AME2012}.
The slopes of the trends for both elements in the region $34 \le N \le 40$ do not change significantly.
On the other hand for two-neutron separation energy lines calculated with the full $pf$-shell phenomenological
GXPF1B5~\cite{HO-EPJA05} interactions,   changes in the slope are observed between $N=35$ and $N=36$ for potassium and calcium isotopes
near the same neutron number where the cross section systematics change.
Because $^{55}$Ca ($N$=35) is predicted  by the shell-model with the GXPF1B5 effective interaction~\cite{GXPF1B}  to have a low one-neutron separation energy  of 0.75~MeV, we might expect that the change of slope in Fig.~\ref{Fig_Qg_all} would correspond to $N=35$ instead of $N=36$.
Thus, it is possible that the trends seen in our cross section data reflect the $N=34$ subshell closure predicted by the GXPF1B5 shell model~\cite{GXPF1B}.
Similar breaks in the slopes are seen in the data at $N=28$, but they are not as dramatic as at $N=34$.

%---------------------------------------------------------------------------
\subsection{Global trends of cross sections\label{secSn}}
A way to visualize the possible effect  of the $N=34$ subshell closure is to
plot the production cross sections versus atomic number. Fig.~\ref{Fig_triton_plot} shows the data connected by lines of constant $N-2Z$, which represent changes between nuclei different by a triton, and the label at each point is the neutron number ($N$).  This ensemble of lines exhibits a large dip at the shell closure at $Z=$~20 in the region of $^{54-56}$Ca highlighted by the red-dashed quadrangle. The same dip can be observed in a similar figure generated from the cross sections measured in the $^{76}$Ge measurements~\cite{OT-PRC09}.

A reason that these trends may be visible in lines of $N-2Z$  in  Fig.~\ref{Fig_triton_plot} is that such nuclei
have approximately the same neutron separation energy and  the drip-line lies close to an $N-2Z$ line for $16 \le Z \le 23$. Further,
each line connects nuclei with either an odd or even number of neutrons; hence, the large odd-even difference in nuclear binding due to pairing is not present along each line.
 For a constant separation energy we expect the cross section to fall smoothly with $Z$ in Fig.~\ref{Fig_triton_plot}.  At $Z=20$, $N=34$ the (sub)shell closures correspond to a lower $S_{1n}$ and $S_{2n}$ and might be responsible for the dip in  the trends, highlighted by the red box.

To illustrate this, two-neutron separation energy $S_{2n}$  versus neutron number ($N$) of elements $12 \le Z \le 22$ calculated with the full $pf$-shell
GXPF1B5 interactions~\cite{GXPF1B} is shown in Fig.~\ref{Fig_s2n_Zconnect}.   The label at each point is the atomic number ($Z$) and the red-dashed quadrangle encompasses the same region as the one in Fig.~\ref{Fig_triton_plot} for isotopes with $34 \le N \le 37$ and $18 \le Z \le 21$.
The $N=34$ subshell closure results in a lowering  of the $S_{2n}$ values for $Z=20$ as seen in the figure, corresponding to the same nuclei that have relatively low production cross sections compared to the $N-2Z$ trends  in Fig.~\ref{Fig_triton_plot}.

Based on the observations discussed above, it seems possible that the  production cross section systematics provide a hint of a change of the nuclear mass surface. The effect is most pronounced close to the drip-line.  Plotting the cross section of elements against $Q_{\text{g}}$ can exhibit sudden changes in slope that are correlated with regions of changes in the nuclear structure, such as (sub)shell closures.

%WWWWWWWWWWWWWWWWWWWWWWWWWWWWWWWWWWWWWWWWWWWWWWWWWWWWWWWWWWWWWWWWWWWWWWWWWWWWWW

\subsection{Abrasion-Ablation model\label{AA}}

In order to test the relationship between cross sections and separation energies, the production cross sections obtained in this experiment were compared with those from calculations with the Abrasion-Ablation model~\cite{OT-NIMB03} implemented in the \lisepp code~\cite{OT-NIMB08}. Results for neutron-rich isotopes of potassium and calcium  using different mass models are shown in Fig.~\ref{Fig_CS_Z19_20}.
Predictions of the \epaxthree systematics are shown on the plots for comparison.
An excitation energy of 15~MeV per abraded nucleon was deduced from an experimental data fit of the Abrasion-Ablation (AA) model with AME2012 masses~\cite{AME2012} and masses deduced from the shell model using the GXPF1B interactions~\cite{HO-EPJA05},  whereas a value 18~MeV has been obtained  with KTYU masses~\cite{KTUY-PTP05}.  The  \lisepp LDM1 paramaterization was used to extrapolate masses of very neutron-rich nuclei absent in  AME2012 and GXPF1B mass predictions. The AA model with GXPF1B masses significantly underestimates cross sections for isotopes with $N \ge 35$ such as $^{54,55}$K and $^{55-57}$Ca, whereas fair agreement is observed  using GXPF1B5 masses. Decreasing  the effective energy gap between adjacent neutron single-particle orbitals $f_{5/2}$ and $p_{1/2}$ in GXPF1B makes isotopes  with $36 \le N \le 40$  around calcium more particle bound, therefore the result of the $^{54}$Ca  $E_x(2_1^+)$  measurement~\cite{RIKEN-54Ca}
is in good agreement with the measured cross sections and whose calculated using the AA model with GXPF1B5 masses.

On the other hand the mass models that don't predict the slope changes at $N=36$ in the $S_{2n}$ figure (see Fig.~\ref{Fig_Sn_All_models}) overestimate the
cross sections for neutron-rich potassium and calcium isotopes (see AA calculations with masses from the KTYU model~\cite{KTUY-PTP05} in Fig.~\ref{Fig_CS_Z19_20}).

%WWWWWWWWWWWWWWWWWWWWWWWWWWWWWWWWWWWWWWWWWWWWWWWWWWWWWWWWWWWWWWWWWWWWWWWWWWWWWW

\subsection{Estimation of $^{60}$Ca production cross section\label{Chapter-Ca60}}

\begin{table}[h]
\caption{Calculated $^{59,60}$Ca production cross section in $^{76}$Ge, $^{82}$Se + $^9$Be reactions. Masses calculated with the GXPF1B5 interaction have been used for the $Q_g$-systematics and the \lisepp Abrasion-Ablation model calculations.\\}\label{Tab_60Ca}
\begin{tabular}{|c|c|c|c|}
\hline
  ~Primary~  & ~Estimation~ &     $^{59}$Ca & $^{60}$Ca \\
  beam       & method       &     ~cross section~ & ~cross section~  \\
             &         &     (mb) & (mb)  \\

\hline
         $^{82}$Se  &       AA           &   1.39e-12  &   4.83e-13\\
         $^{82}$Se  &       $Q_g$        &   1.62e-12 &    3.19e-13\\
         $^{76}$Ge  &       $Q_g$        &   1.36e-12 &   2.47e-13\\
         $^{82}$Se  &       \epaxthree   &   3.73e-13 &   1.65e-14\\
         $^{76}$Ge  &       \epaxthree   &   3.68e-13 &   1.71e-14\\

\hline
\end{tabular}
\end{table}

Based on experimental $Q_{\text{g}}$-systematics~(Fig.\ref{Fig_Qg_all}) and Abrasion-Ablation model calculations~(Fig.~\ref{Fig_CS_Z19_20}) using GXPF1B5 masses, it is possible to estimate the production cross sections for the next unobserved calcium isotopes using a $^{82}$Se beam (See Table~\ref{Tab_60Ca}). According to this extrapolation, to observe one event of  $^{59}$Ca, the beam intensity has to be increased by a factor of 2 compared to this experiment.  For the case of $^{60}$Ca, at least an order magnitude higher beam intensity is needed. It is important to note, that the \epaxthree systematics~\cite{KS-EPAX3} predict a factor of 20 less production for the $^{60}$Ca isotope, compared to both $Q_{\text{g}}$-systematics and AA calculations ($4(\pm{1})\times 10^{-13}$~mb), that makes the search for $^{60}$Ca more likely to be successful in the near future.

%WWWWWWWWWWWWWWWWWWWWWWWWWWWWWWWWWWWWWWWWWWWWWWWWWWWWWWWWWWWWWWWWWWWWWWWWWWWWWW

\section{Summary\label{Summary}}

The present study of fragmentation of a $^{82}$Se beam at
139~MeV/u found evidence for  four previously
unobserved neutron-rich isotopes ($^{64}$Ti, $^{67}$V, $^{69}$Cr,
$^{72}$Mn). The longitudinal momentum
distributions and cross sections for a large number of neutron-rich
nuclei produced by the $^{82}$Se beam were measured by varying the
target thickness in a two-stage fragment separator using a narrow momentum selection.
The momentum distributions of 126
neutron-rich isotopes of elements with \protect{$11\le Z\le 32$}
were compared to models that describe the shape and centroid of
momentum distributions.  From these measurements we have obtained a new set of parameters
for the semiempirical momentum distribution models~\cite{DJM-PRC89,AG-PLB74}.

The most neutron-rich nuclei of elements with $Z=$~19 to 21 have been produced with an enhanced rate compared to the systematics
of the production cross sections from the $Q_g$ systematics.
This trend was previously reported for the fragmentation of a $^{76}$Ge beam~\cite{OT-PRL09},
and therefore the current results confirm those of our previous experiment.
This is an indication  of a change in the nuclear mass surface near $Z = 20$ for very neutron-rich nuclei.
This result has been explained with a  shell model that predicts a subshell closure at $N=$~34 and a more pronounced one at $Z=$~20.
We have shown that production cross section systematics can provide a hint of a change of the nuclear mass surface close to the drip-line.
 Plotting the cross section of elements against $Q_{\text{g}}$ can exhibit sudden changes in slope that are correlated with regions of changes in the nuclear structure, such as (sub)shell closures.  A correlation to the nuclear mass models has been shown via plots of the two-neutron separation energy.

It has been shown that the Abrasion-Ablation model is very sensitive to the input mass values for the most exotic nuclei.
The measured cross sections were best reproduced by using masses derived from the full {\it pf} shell model space with the GXPF1B5~\cite{HO-EPJA05}
effective interaction modified to a recent $^{54}$Ca $E_x(2_1^+)$ measurement~\cite{RIKEN-54Ca}.

The cross section for production of $^{60}$Ca using a $^{82}$Se beam on beryllium has been estimated at $4(\pm{1})\times 10^{-16}$~barn.
This estimate is based on both $Q_{\text{g}}$-systematics and Abrasion-Ablation calculations using masses derived from the shell-model effective interaction GXPF1B5~\cite{GXPF1B}.

%WWWWWWWWWWWWWWWWWWWWWWWWWWWWWWWWWWWWWWWWWWWWWWWWWWWWWWWWWWWWWWWWWWWWWWWWWWWWWW
\begin{acknowledgments}
The authors would like to acknowledge the operations staff of the
NSCL for developing the intense $^{82}$Se beam necessary for this
study. This work was supported by the U.S.~National Science
Foundation under grants PHY-06-06007, PHY-10-68217, and PHY-11-02511.
Discussions with Prof.~V.~G.~Zelevinsky are very appreciated.

\end{acknowledgments}

\bibliography{82Se}

%WWWWWWWWWWWWWWWWWWWWWWWWWWWWWWWWWWWWWWWWWWWWWWWWWWWWWWWWWWWWWWWWWWWWWWWWWWWWWW

%
% ****** End of file 82Se-PRC.tex *****
 \end{document}